\documentclass[11pt, a4paper]{article}

\usepackage{enumerate,appendix}
\usepackage{braket}
\usepackage{amsmath}
\usepackage{yhmath}
\usepackage{amsfonts}
\usepackage{amssymb}
\usepackage{mathrsfs}
\usepackage{physics}
\usepackage{hyperref}
\usepackage{graphicx, rotating}
\usepackage{epstopdf}
\usepackage{epsfig}
\usepackage{latexsym}
\usepackage{color}
\usepackage{cite}
\usepackage{slashed}
\usepackage{dsfont} 
\usepackage{mathtools}
\usepackage{bbold}
\usepackage{cancel}
\usepackage[normalem]{ulem} 
\usepackage{bm}
\usepackage[dvipsnames]{xcolor}
\usepackage{comment}
\usepackage[utf8]{inputenc}
\usepackage{soul}
\usepackage{multirow}
\usepackage{graphics,subfigure}
\usepackage[]{breqn} 

\hypersetup{colorlinks, citecolor=bluscuro, linkcolor=black, urlcolor=bluscuro}
\definecolor{rossos}{cmyk}{0,1,1,0.55}
\definecolor{bluscuro}{rgb}{0.15, 0.2, .85}
\definecolor{bluchiaro}{cmyk}{1,.3,0.,0.1}

\newcommand{\be}{\begin{equation}}
\newcommand{\ee}{\end{equation}}
\newcommand{\bea}{\begin{eqnarray}}
\newcommand{\eea}{\end{eqnarray}}

\newcommand{\orcid}{\includegraphics{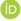}}
\newcommand{\orcidlink}[1]{\href{https://orcid.org/#1}{{\orcid}}}

\begin{document}

\begin{titlepage}
\begin{flushright}
{IFT-UAM/CSIC-25-27}
\end{flushright}
\begin{center} ~~\\
\vspace{0.5cm} 
\Large {{\bf\Large
 Practical Criteria for Entanglement and Nonlocality in Systems with Additive Observables}
} 
\vspace*{1.2cm}

\normalsize{
{\bf 
Alexander Bernal\footnote{
alexander.bernal@csic.es}$^a$\orcidlink{0000-0003-3371-5320}
,
J.~Alberto Casas\footnote{j.alberto.casas@gmail.com
}$^a$\orcidlink{0000-0001-5538-1398}
 and
Juan Falceto\footnote{falcetoj@tcd.ie}$^{ab}$\orcidlink{0009-0005-4846-1329}

} 
 
\smallskip  \medskip
$^{a}${\it Instituto de F\'\i sica Te\'orica, IFT-UAM/CSIC,}\\
\it{Universidad Aut\'onoma de Madrid, Cantoblanco, 28049 Madrid, Spain\\ \hspace{1cm} 
\\
$^{b}$School of Mathematics \& Hamilton Mathematics Institute,\\
Trinity College Dublin, Ireland}}

\medskip

\vskip0.6in 

\end{center}

\centerline{ \large\bf Abstract }
\vspace{.5cm}

For general bipartite mixed states, a sufficient and necessary mathematical condition for certifying entanglement and/or (Bell) non-locality remains unknown. In this paper, we examine this question for a broad and physically relevant class of bipartite systems, specifically those possessing an additive observable with a definite value. 
Such systems include, for example, final states of particle decays or bipartitions of spin chains with well-defined magnetization. We derive very simple, handy criteria for detecting entanglement or non-locality in many cases. For instance, if $\rho_{\left( m  p\right)\left( nq\right)}
\neq 0$, where the eigenstates $\ket{np}$ or $\ket{mq}$ do not correspond to the given definite value of the additive observable, then the state is necessarily entangled—this condition is very easy to check in practice. If, in addition, the partitioned Hilbert space has dimension $2\times d$, the condition becomes necessary. Furthermore, if the sectors associated with the eigenstates $\ket{mp}$ or $\ket{nq}$ are non-degenerate, there exists a CHSH inequality that is violated. We illustrate these results by analyzing the potential detection of entanglement and nonlocality  in Higgs $\rightarrow ZZ$ decays at the LHC.

\vspace*{2mm}
\end{titlepage}


\tableofcontents

\section{Introduction}
\label{sec:Intro}

For pure states mathematical criteria for both entanglement and (Bell) nonlocality are somehow trivial. Namely, a state $\ket{\psi}\in {\cal H}_A\otimes{\cal H}_B$ is entangled iff the reduced density matrix, $\rho_A= {\rm Tr}_B\ \rho$, with $\rho=\ketbra{\psi}$, corresponds to a mixed state (tracing over $A$ indices leads to the same result).
Hence, a necessary and sufficient condition for entanglement is ${\rm Tr}\ \rho_A^2\neq 1$ or, equivalently a non-vanishing von Neumann entropy, $S_{\rm vN}=-{\rm Tr}\ \rho_A\ln \rho_A\neq 0$. Concerning Bell nonlocality, it was shown long time ago that for pure states entanglement and nonlocality are equivalent concepts. More precisely, given an entangled pure state of a bipartite system, one can always construct a CHSH inequality that is violated \cite{GISIN1991201,Popescu:2002won}.

In contrast, for general mixed states these questions are much harder and remain to a great extent unresolved. There is not a general sufficient and necessary mathematical condition to certify entanglement. The celebrated Peres-Horodecki criterion \cite{PhysRevLett.77.1413} offers a useful sufficient condition for entanglement, namely that the partially transposed density matrix, $\rho^{T_2}$, contains negative eigenvalues. However this does not represent a necessary condition beyond the low-dimensional cases,  qubit-qubit or qubit-qutrit \cite{Horodecki:1996nc}. Regarding nonlocality, for mixed states entanglement is itself a necessary but not sufficient condition for Bell-violation \cite{PhysRevA.40.4277}. In the case of qubit-qubit systems, the maximal Bell violation (in measurements involving two observables for both, Alice and Bob) is provided by a CHSH inequality. The corresponding optimal observables, $A_1, A_2, B_1, B_2$ , and the amount of violation is given by a rather simple recipe \cite{HORODECKI1995340}. However, beyond this lowest-dimension case there is no similar strategy to certify Bell-violation (see, however, the result on qubit-qudit systems presented in \cite{Bernal:2024fio}).

Still, in certain physically relevant scenarios one can find general rules. As was shown in ref. \cite{Aguilar-Saavedra:2022wam,Bernal:2023ruk,Bernal:2024xhm}, for Higgs decays into two massive gauge bosons ($ZZ$ or $WW$), conservation of angular momentum along the decay direction implies a particular texture of the density matrix, which in turn implies that the Peres-Horodecki criterion is a sufficient {\em and} necessary condition, despite being a system of two qutrits. One can wonder whether this feature is a peculiar characteristic 
of this particular system, or it reflects a general property of a much larger class. Besides, one can speculate about the situation of Bell violation in such systems, exploring in particular the  existence of a recipe to certify nonlocality.

In this paper we will consider a quite general and physically relevant class of (bipartite) systems. Namely, those for which there exists an additive observable, say $\hat J=\hat J_A+\hat J_B$, which has a definite value, say $J$. This is certainly the case for the above-mentioned Higgs decays, with $\hat J\equiv J_z$, i.e. the angular-momentum component along the direction of decay in the C.M. frame. Other systems with the same characteristic are meson decays; spin chains under a given bipartition, where the role of $\hat J$ could be played by e.g. the magnetization; and hybrid (discrete-continuous) systems, such as atom-cavity setups, where the role of $\hat{J}$ could be played by the energy or by some parity-like property. Of course, the details will depend on the physical scenario and the particular state under consideration, but the common characteristic is the existence of such additive observable with a well-defined value. 

Interestingly, this kind of setup arises in a natural way whenever the symmetries of the Hamiltonian lead to the existence of a conserved quantity. In that instance, if the initial state has a well-defined value of such quantity the system will remain in the same class  along unitary evolution. This is the case of the above-mentioned examples, such as Higgs boson decays.

In Section \ref{sec:setup} we describe the setup and establish the notation. Section \ref{sec:Entanglement} is devoted to the study of entanglement in this kind of systems, providing simple recipes to certify it in many cases. In Section \ref{sec:CHSH} we perform a similar analysis for the nonlocality in these systems. In Section \ref{sec:Bell_higgs} we apply the previous results to the study of the
$H\rightarrow ZZ$ system.
Finally, Section \ref{sec:conclusions}  summarizes our outcomes and conclusions.

\section{The setup}
\label{sec:setup}

We start by defining the setup considered throughout the paper and introducing our notation. We consider a bipartite system with Hilbert space ${\cal H}={\cal H}_A\otimes {\cal H}_B$ (Alice and Bob) of unrestricted finite dimensions $\mathrm{dim}(\mathcal{H}_A) = d_A$, $\mathrm{dim}(\mathcal{H}_B) = d_B$ and an additive observable
\bea
\hat J=\hat J_A+\hat J_B,
\eea
where $\hat J_A=\hat J_A\otimes\mathds{1}_B$, $\hat J_B=\mathds{1}_A\otimes\hat J_B$ is understood. The observable $\hat J$ does not need to possess a particular physical meaning. Our only requirement is that the state under consideration (whether pure or mixed) has a well-defined value of $\hat J$, say $J$. This is equivalent to the condition $\Pi_J\, \rho\, \Pi_J =\rho$, where $\Pi_J$ is the projector onto the subspace of $\cal H$ associated with the eigenvalue $J$. 

For the Alice's (Bob's) Hilbert space we will use a basis of eigenstates of $\hat J_A$ ($\hat J_B$), with the following notation:
\begin{equation}
\begin{aligned}\label{eq:basis}
    {\cal H}_A:&\ \ \{|{m}\rangle\},\ \ {\rm s.t.}\ \ \hat J_A |{m}\rangle = M |{m}\rangle, \\
{\cal H}_B:&\ \ \{\, \ket{p}\, \},\ \ {\rm s.t.}\ \ \hat J_B \ket{p} = P \ket{p}.
\end{aligned}
\end{equation}

In general we will use $m,n$ ($p,q$) for Alice's (Bob's) indices. The corresponding eigenvalues
can be degenerate, so generically there can be several $\ket{m}$ ($\ket{p}$) states for each eigenvalue $M$ ($P$) of $\hat J_A$ ($\hat J_B$). 

The basis of the global Hilbert space ${\cal H}$ is the tensor product of the previous ones, $\{\ket{m p}=\ket{m}\otimes\ket{p}\}$. The corresponding eigenvalues of $\hat J$ are $M+P$. The assumption that $\hat J$ has a well-defined value, $J$, implies that only basis states $\ket{m p}$ with 
\bea
M+P=J
\eea
can be represented in the density matrix. In other words,
\bea
\rho_{\left( m  p\right)\left(n q\right)}\neq 0\ \Rightarrow\ M+P=N+Q=J.
\label{nonvancond}
\eea
Note that this additivity condition does not fix in general the values of $M$ and $P$ (and $N$ and $Q$), since there can be many combinations of them that fulfill it.

\section{Entanglement}
\label{sec:Entanglement}

\subsection{The structure of $\rho$ and $\rho^{T_2}$}
 
As it is well known, the partially transposed matrix,
\bea
\rho^{T_2}_{\left( m  q\right)\left( n p\right)}=\rho_{\left( m  p\right)\left(n q\right)}
\label{rhorhoT2}
\eea
plays a central role in the Peres-Horodecki criterion for entanglement; namely, if $\rho^{T_2}$ has some negative eigenvalue, the state is necessarily entangled. 

Let us first show that in the setup described in the previous section (i.e. a system with a well-defined value of an additive observable) 
the $\rho^{T_2}$ matrix is block-diagonal, with two different types of blocks.

\vspace{0.2cm}
\noindent
Consider a diagonal element of $\rho^{T_2}$ (and thus of $\rho$),
$
\rho^{T_2}_{\left( m  q\right)\left( m q\right)}=\rho_{\left( m  q\right)\left(m q\right)}
$.
Denoting $M, Q$ the eigenvalues of $\hat J_A, \hat J_B$ associated with $\ket{m}, \ket{q}$, there are two relevant possibilities: $M+Q=J$ or $M+Q\neq J$. According to the additivity condition (\ref{nonvancond}), only in the first case the diagonal element can be non-vanishing,
\bea
\rho^{T_2}_{\left( m  q\right)\left( m q\right)}=\rho_{\left( m  q\right)\left(m q\right)}
\left\{\begin{array}{cc}
 \neq 0 & \ \ \text{only if $M+Q=J$},
  \\[6mm]
	=0  & \ \ \text{if $M+Q\neq J$}.
	\end{array}\right.
\label{two cases}	
\eea
Let us start with the first case, $M+Q=J$. Consider an element in the row of $\rho^{T_2}$ whose associated diagonal element belongs to this class, say $\rho^{T_2}_{\left( m  q\right)\left( n p\right)}=\rho_{\left( m  p\right)\left(n q\right)}$. From Eq.(\ref{nonvancond}), such element can only be non-vanishing if $M+P=N+Q=J$, which in turn implies $N=M$ and $P=Q$. 

Denote $(mq)_\alpha$ the $(m,q)$ pairs associated with the eigenvalues $M,Q$; so $\alpha=1,\ \dots, {\rm deg}\,M\cdot {\rm deg}\,Q$, where ${\rm deg}\,M$ (${\rm deg}\,Q$) is the degeneracy of $M$ ($Q$). Then any non-vanishing element in the row must be of the form $\rho^{T_2}_{\left( m  q\right)_\alpha\left( m q\right)_{\alpha'}}$. 
Consequently, all the (possibly) non-vanishing entries of a row of $\rho^{T_2}$ where the first index is $(m,q)$, with $M,Q$ fixed and $M+Q=J$, form a principal submatrix of $\rho^{T_2}$ of dimension $({\rm deg}\,M\cdot {\rm deg}\,Q)\times ({\rm deg}\,M\cdot {\rm deg}\,Q)$,
\begin{equation} 
    \begin{pmatrix}
     \rho^{T_2}_{\left(m q\right)_1\left(m q\right)_1} & \rho^{T_2}_{\left(m q\right)_1\left(m q\right)_2} & \rho^{T_2}_{\left(m q\right)_1\left(m q\right)_3} & \hdots \\
     \rho^{T_2}_{\left(m q\right)_2\left(m q\right)_1}  & \rho^{T_2}_{\left(m q\right)_2\left(m q\right)_2} & \rho^{T_2}_{\left(m q\right)_2\left(m q\right)_3} & \hdots \\
     \rho^{T_2}_{\left(m q\right)_3\left(m q\right)_1}  & \rho^{T_2}_{\left(m q\right)_3\left(m q\right)_2} & \rho^{T_2}_{\left(m q\right)_3\left(m q\right)_3} & \hdots \\
     \vdots  & \vdots & \vdots & \ddots  
    \end{pmatrix}.
    \label{eq:rho_tracefull}
\end{equation}

\vspace{0.2cm}
Let us consider now the second case in Eq.(\ref{two cases}), $M+Q\neq J$, so that the diagonal element $\rho^{T_2}_{\left( m  q\right)\left( mq\right)}=\rho_{\left( m  q\right)\left(m q\right)}$ is vanishing. An element in the row of this entry of $\rho^{T_2}$, say $\rho^{T_2}_{\left( m  q\right)\left( n p\right)}=\rho_{\left( m  p\right)\left(n q\right)}$, can only be non-vanishing if $M+P=N+Q=J$, which in turn implies $N=J-Q$, $P=J-M$ (both unique). Note also that  $N\neq M$, $P\neq Q$ and $N+P=2J-(M+Q)\neq J$. Similarly to the previous case, we denote by $(mq)_\beta$, $(np)_\gamma$ the pairs $(m, q)$, $(n,p)$ associated with the eigenvalues $(M,Q)$, $(N,P)$, respectively. Then, all the (possibly) non-vanishing entries of a row of $\rho^{T_2}$ where the first index is $(m,q)_\beta$,  
or $(n,p)_\gamma$ 
belong to a principal submatrix of $\rho^{T_2}$ of dimension $({\rm deg}\,M\cdot {\rm deg}\,Q + {\rm deg}\,N\cdot {\rm deg}\,P)\times ({\rm deg}\,M\cdot {\rm deg}\,Q+ {\rm deg}\,N\cdot {\rm deg}\,P)$,
\begin{equation}
\begin{multlined}
    \left(\begin{array}{ccc|ccc} 
     \rho^{T_2}_{\left(  m   q\right)_1\left(  m   q\right)_1} & \rho^{T_2}_{\left(  m   q\right)_1\left(  m   q\right)_2} & \hdots & \rho^{T_2}_{\left(  m   q\right)_1\left(  n   p\right)_1} &  \rho^{T_2}_{\left(  m   q\right)_1\left(  n   p\right)_2} & \hdots \\
     \rho^{T_2}_{\left(  m   q\right)_2\left(  m   q\right)_1}  & \rho^{T_2}_{\left(  m   q\right)_2\left(  m   q\right)_2} & \hdots & \rho^{T_2}_{\left(  m   q\right)_2\left(  n   p\right)_1} & \rho^{T_2}_{\left(  m   q\right)_2\left(  n   p\right)_2}& \hdots \\  
     \vdots &    \vdots &   \ddots &   \vdots &   \vdots & \ddots \\
     \hline
     \rho^{T_2}_{\left(  n   p\right)_1\left(  m   q\right)_1}  & \rho^{T_2}_{\left(  n   p\right)_1\left(  m   q\right)_2} & \hdots & \rho^{T_2}_{\left(  n   p\right)_1\left(  n   p\right)_1} & \rho^{T_2}_{\left(  n   p\right)_1\left(  n   p\right)_2}& \hdots \\
     \rho^{T_2}_{\left(  n   p\right)_2\left(  m   q\right)_1}  & \rho^{T_2}_{\left(  n   p\right)_2\left(  m   q\right)_2} & \hdots & \rho^{T_2}_{\left(  n   p\right)_2\left(  n   p\right)_1} & \rho^{T_2}_{\left(  n   p\right)_2\left(  n   p\right)_2}& \hdots \\
     \vdots  &  \vdots  &  \ddots  &  \vdots  & \vdots &  \ddots 
\end{array}\right).
\end{multlined}
\end{equation}
Now, the two diagonal sub-blocks of this matrix are vanishing since their entries (which are related to $\rho$ entries by (\ref{rhorhoT2})) do not satisfy the additivity condition (\ref{nonvancond}),
so this principal matrix reads
\begin{equation}
\begin{multlined}
\left(\begin{array}{ccc|ccc} 
     0 & 0 & \hdots & \rho^{T_2}_{\left(  m   q\right)_1\left(  n   p\right)_1} &  \rho^{T_2}_{\left(  m   q\right)_1\left(  n   p\right)_2} & \hdots \\
     0  & 0 & \hdots & \rho^{T_2}_{\left(  m   q\right)_2\left(  n   p\right)_1} & \rho^{T_2}_{\left(  m   q\right)_2\left(  n   p\right)_2}& \hdots \\  
     \vdots &    \vdots &   \ddots &   \vdots &   \vdots & \ddots \\ 
     \hline
     \rho^{T_2}_{\left(  n   p\right)_1\left(  m   q\right)_1}  & \rho^{T_2}_{\left(  n   p\right)_1\left(  m   q\right)_2} & \hdots & 0 & 0 & \hdots \\
     \rho^{T_2}_{\left(  n   p\right)_2\left(  m   q\right)_1}  & \rho^{T_2}_{\left(  n   p\right)_2\left(  m   q\right)_2} & \hdots & 0 & 0& \hdots \\
     \vdots  &  \vdots  &  \ddots  &  \vdots  & \vdots &  \ddots
\end{array}\right).
    \label{eq:rho_traceless}
\end{multlined}
\end{equation}
Since this is a traceless Hermitian matrix, if any off-diagonal element is different from zero, it necessarily contains a negative eigenvalue. In that case the Peres-Horodecki criterion applies and the global state is entangled. We call an element of this kind a ``crossed off-diagonal entry".
In other words, if the density matrix $\rho$ contains a non-vanishing crossed off-diagonal entry,
\be
\rho_{\left( m  p\right)\left( nq\right)}
\neq 0,\ \ \ {\rm with}\ M+Q\neq J\ ,
\label{crossed}
\ee
then the state is necessarily entangled by the Peres-Horodecki criterion. Note that this result is very handful in practice since it allows identification of entanglement without the need to construct the $\rho^{T_2}$ matrix and evaluate its eigenvalues, which can be a lengthy procedure when the dimensions of the Hilbert spaces are large.

\subsection{Implications for entanglement}

We have just seen that, for a system with a well-defined value of an additive observable, the presence of a non-vanishing crossed off-diagonal entry (\ref{crossed}) in the density matrix implies that the state is entangled and, besides, it satisfies the Peres-Horodecki criterion. 

When {\em all} crossed-off diagonal entries are vanishing, the state may still be entangled thanks to the off-diagonal entries of the blocks  of $\rho^{T_2}$ associated with $M, Q$ sectors for which $M+Q=J$, Eq.(\ref{eq:rho_tracefull}). 
Consider one of such sectors.
Following the notation previously introduced,
let us call $( m  q)_\alpha$, with $\alpha=1,\ \dots, {\rm deg}\,M\cdot {\rm deg}\,Q$, the pairs associated with the eigenvalues $M,Q$. The corresponding block in $\rho^{T_2}$ has the form (\ref{eq:rho_tracefull}), which we reproduce here for convenience:
\begin{equation} 
    \begin{pmatrix}
     \rho^{T_2}_{\left(m q\right)_1\left(m q\right)_1} & \rho^{T_2}_{\left(m q\right)_1\left(m q\right)_2} & \rho^{T_2}_{\left(m q\right)_1\left(m q\right)_3} & \hdots \\
     \rho^{T_2}_{\left(m q\right)_2\left(m q\right)_1}  & \rho^{T_2}_{\left(m q\right)_2\left(m q\right)_2} & \rho^{T_2}_{\left(m q\right)_2\left(m q\right)_3} & \hdots \\
     \rho^{T_2}_{\left(m q\right)_3\left(m q\right)_1}  & \rho^{T_2}_{\left(m q\right)_3\left(m q\right)_2} & \rho^{T_2}_{\left(m q\right)_3\left(m q\right)_3} & \hdots \\
     \vdots  & \vdots & \vdots & \ddots  
    \end{pmatrix}.
    \label{eq:rho_tracefull2}
\end{equation}
This submatrix of $\rho^{T_2}$ arises, by partial transposition of Bob's indices, from a principal submatrix of $\rho$. To see this, note that $\rho= ({\rho^{T_2}})^{T_2}$. This partial transposition amounts to interchange the Bob's indices, 
(the $q$'s), so the elements obtained in this way still belong to the block (\ref{eq:rho_tracefull2}). Hence, under partial transposition, the submatrix (\ref{eq:rho_tracefull2}) only undergoes a reordering of the entries, but it is still a square principal submatrix, and in particular positive semi-definite.

Consequently if all the crossed off-diagonal entries are zero, as assumed, then all non-diagonal elements of both $\rho$ and $\rho^{T_2}$ belong to blocks as the one depicted above. So $\rho$ is a block diagonal matrix too, 
with a block for each $(M, Q)$ sector of the Hilbert space with $M+Q=J$. \footnote{Note that $\rho$ can be written as a convex sum of density matrices
$\rho=\sum p_{MQ}\, \rho_{MQ}$, where each  $p_{MQ}\, \rho_{MQ}$ term corresponds to a block.}
This reduces the task to examine if these blocks are entangled or not. If all of them are separable, then the global state is separable. Otherwise it is entangled. It turns out that the structure of the blocks is greatly determined by the degeneracy of the corresponding eigenvalues, $J_A=M$ and $J_B=Q$. We distinguish here several possibilities:

\begin{description}
    \item [Non-degenerate case]
    \hspace{1cm}

If the two eigenvalues $M,Q$ defining the sector are both non-degenerate, the submatrix (\ref{eq:rho_tracefull2}) is formed of a single (and diagonal) element, so the sector is obviously separable. Notice that if all the $M,Q$ sectors are non-degenerate, all non-vanishing off-diagonal entries in $\rho$ correspond to crossed off-diagonal entries,  Eq.~(\ref{crossed}).  Then the presence of an off-diagonal entry in the density matrix $\rho$ guarantees that the state is entangled (and the Peres-Horodecki criterion is sufficient and necessary). This is the case of particle decays into two particles, e.g. $H\rightarrow ZZ$. 

\item [One non-degenerate eigenvalue]
\hspace{1cm}
   
Suppose that one of the eigenvalues of the sector, say $Q$, is non-degenerate. Then there is only one state $|{q}\rangle\in {\cal H}_B $ involved, so the sector is separable. Incidentally, the corresponding submatrix of $\rho^{T_2}$, Eq.~(\ref{eq:rho_tracefull2}), is invariant under $T_2$-transposition, so it is also a principal submatrix of $\rho$ and thus positive semi-definite. Hence, the Peres-Horodecki criterion is trivially fulfilled. A direct application of this result is for qubit-qudit systems. Then all sectors have non-degenerate Alice's eigenvalue (unless $\hat{J}_A\propto\mathds{1}_2 $, i.e. trivial), and consequently the only source of entanglement are crossed off-diagonal terms, and the Peres-Horodecki criterion holds as a sufficient {\em and} necessary condition.

\item [Two degenerate eigenvalues]
\hspace{1cm}

When the two eigenvalues of the sector under consideration are degenerate, we can apply the Peres-Horodecki criterion to the corresponding block; i.e. if the submatrix (\ref{eq:rho_tracefull2}) has at least one negative eigenvalue, the sector is entangled. Now, in general the Peres-Horodecki criterion for entanglement is sufficient but not necessary.
Still, if the corresponding degenerate subspaces are of dimension $2\otimes2$ or $2\otimes3$ (or $3\otimes2$) the Peres-Horodecki criterion is also necessary \cite{Horodecki:1996nc}.
In other words, if the degeneracy of $(M,Q)$ is at most $(2,3)$ (or $(3,2)$), then the Peres-Horodecki criterion for entanglement is sufficient {\em and} necessary. Otherwise, it remains as a sufficient condition.
\end{description}

\subsection{Summary}
\label{subsec:summary_ent}

We summarize the previous results concerning entanglement for a bipartite system when there is
an additive observable $\hat J=\hat J_A + \hat J_B$ with a well-defined value, $J$. First, if the density matrix contains a non-vanishing crossed off-diagonal entry,
$
\rho_{\left( m  p\right)\left( nq\right)}
\neq 0$ with $M+Q\neq J$,
then the state is necessarily entangled and satisfies the Peres-Horodecki criterion.

If all the crossed off-diagonal entries are vanishing, 
the Peres-Horodecki criterion is a sufficient {\em and} necessary condition provided that for any pair of eigenvalues $\left(M,Q\right)$ satisfying $M+Q=J$, the degeneracy ${\rm deg}\,M\times {\rm deg}\,Q$ is either
\begin{enumerate}
    \item ${\rm deg}\,M\times 1\ $ or $\ 1\times {\rm deg}\,Q$.
    \item $2\times2$,~ $2\times3$~ or~ $3\times2$.
\end{enumerate}
Furthermore, if for all pairs $\left(M,Q\right)$ the degeneracy is of type 1, then $\rho$ is entangled if and only if there exists a nonzero crossed off-diagonal element, Eq.~(\ref{crossed}). Consequently, the Peres-Horodecki criterion holds as a sufficient {\em and} necessary condition. This is the case of qubit-qudit systems, where all the sectors have non-degenerate Alice's eigenvalue (unless $\hat{J}_A\propto\mathds{1}_2 $, i.e. trivial).

For the degeneracy of type 2, the presence of crossed off-diagonal elements is a sufficient condition for entanglement, but not necessary. Even in its absence, the entanglement can arise from the degenerate subspaces. 

When the degeneracy of the bipartite system is larger than that of types 1 and 2, then the Peres-Horodecki criterion remains as a sufficient condition for entanglement.
    
\section{Bell nonlocality}
\label{sec:CHSH}

As mentioned in the introduction, section \ref{sec:Intro}, in bipartite quantum systems Bell non-locality (i.e. violation of Bell-like inequalities) is guaranteed for any {\em pure} entangled state \cite{GISIN1991201,Popescu:2002won}; however this is not longer true for general mixed states \cite{PhysRevA.40.4277}. In many cases entanglement does not imply non-locality (although the opposite is always true). Actually, checking the non-locality of a mixed state is not an easy matter. Beyond dimension 3, there is no known set of inequalities whose violation is a necessary and sufficient condition for non-locality.
In a similar spirit to the previous section, we will explore in this section how the existence of a well-defined additive quantity changes this picture. 

\subsection{The CHSH inequality}
Let us briefly review the most well-known Bell-type inequality, the CHSH inequality \cite{Bell:1964kc,Clauser:1969ny}, which will play the central role in the present section. 

As usual, we consider a bipartite quantum system with Hilbert space $\mathcal{H} = \mathcal{H}_A \otimes \mathcal{H}_B$ (Alice and Bob) of dimensions $\mathrm{dim}(\mathcal{H}_A) = d_A$, $\mathrm{dim}(\mathcal{H}_B) = d_B$. 
Besides, let us assume that Alice and Bob measure two observables with 2 distinct outcomes each. Thus, the measurements  of Alice (Bob) are parametrized by $d_A\times d_A$ ($d_B\times d_B$) Hermitian operators $A_1, A_2$ ($B_1,B_2$) with $\pm 1$ eigenvalues under some degeneracy. Consider now the operator, 
\begin{equation}
    A_1\otimes(B_1+B_2) + A_2\otimes(B_1-B_2).
    \label{CHSH1}
\end{equation}
Classically (i.e. if $A_{1,2}$, $B_{1,2}$ have well-defined values), the value of this operator is forced to be $+2$ or $-2$. Hence, its expected value must fulfill
\bea
E\left[A_1\otimes(B_1+B_2) + A_2\otimes(B_1-B_2)\right] \leq 2.
\eea
For a quantum system described by the density matrix $\rho$, the expectation value of an observable $M$ is given by the Born rule $E\left[M\right] = \langle M \rangle_\rho = \Tr{\rho M}$. Consequently, for a quantum bipartite system the CHSH inequality reads
\begin{equation}
\begin{aligned}\label{eq:sum_chsh_popescu}
    F(\rho)=&\left\langle A_1\otimes(B_1+B_2) + A_2\otimes(B_1-B_2) \right\rangle_\rho \\ \bigskip
    =&\Tr{\rho\left[A_1\otimes(B_1+B_2) + A_2\otimes(B_1-B_2) \right]}\leq 2.
\end{aligned}
\end{equation}
Notice that there is one CHSH inequality for each choice of the four observables $A_{1,2}$, $B_{1,2}$, which generically represents a huge parameter space. In order to check non-locality it is enough to find a choice of these observables that violates (\ref{eq:sum_chsh_popescu}). Let us finally mention that the CHSH inequalities are optimal for qubit-qubit systems \cite{Pironio_2014}. Beyond that dimension, there exist other (typically more powerful) inequalities, like the CGLMP ones \cite{Collins:2002sun}, but the broad question of testing nonlocality in general bipartite systems has not been solved. 

\subsection{Violation of the CHSH inequality}
\label{sec:CHSHviol}

As described in section \ref{sec:setup}, let us assume that the bipartite system is in a (generically mixed) state with a well-defined value of an additive observable, $\hat J=\hat J_A+\hat J_B$, say $J$. As in the previous section, we will use the Hilbert-space bases associated with these observables, as given in Eq.
\eqref{eq:basis}.

In this subsection we will prove the following statement: for the quantum system in hand there exists an CHSH inequality (\ref{eq:sum_chsh_popescu}) which is violated provided there is a non-vanishing off-diagonal entry of $\rho$, say 
\begin{equation}
    \rho_{\left( m_0 p_0\right)\left(n_0 q_0\right)}\neq 0\ ,
    \label{elemento}
\end{equation}
such that
\begin{equation}\label{eq:SymAssump}
\begin{aligned}
    M_0+P_0=J,&\quad N_0+Q_0=J,\\
    M_0\neq N_0,&\quad P_0\neq Q_0,\\
    \left(N_0,Q_0\right)&\quad \text{ non-degenerate},
\end{aligned}
\end{equation}
where 
\begin{equation}\label{anchor2}
\begin{aligned}
    \hat{J}_A\ket{m_0} = M_0\ket{m_0}, &\quad \hat{J}_A\ket{n_0} = N_0\ket{n_0}, \\
    \hat{J}_B\ket{p_0} = P_0\ket{p_0}, &\quad \hat{J}_B\ket{q_0} = Q_0\ket{q_0}.
\end{aligned}
\end{equation}
We call $\rho_{\left(m_0 p_0\right)\left(n_0 q_0\right)}$ an {\bf anchor element}.
The conditions in the first line of Eq.~(\ref{eq:SymAssump}) stem from $\Pi_J\, \rho\, \Pi_J =\rho$, and are mandatory for the density matrix element (\ref{elemento}) to be non-vanishing. The condition of the second line corresponds to the definition of a ``crossed off-diagonal entry",
see Eq.\eqref{crossed}.
The last line of (\ref{eq:SymAssump}) is an extra non-degeneracy condition that we impose on the $N_0, Q_0$ eigenvalues. Note that this condition is not imposed on  the whole spectrum of $\hat{J}_A$ and/or $\hat{J}_B$. Notice also that if the non-degenerate eigenvalues are $M_0, P_0$, rather than $N_0, Q_0$, everything goes equivalently since in that case the role of the anchor entry could be played by
$\rho_{\left(n_0 q_0\right)\left( m_0 p_0\right)}=\rho_{\left(m_0 p_0\right)\left( n_0 q_0\right)}^*\neq 0$.

Let us assume that there is indeed an entry $\rho_{\left(m_0 p_0\right)\left( n_0 q_0\right)}\neq 0$ fulfilling Eqs.\eqref{elemento}, \eqref{eq:SymAssump}, which thus serves as an anchor entry. In order to build an appropriate CHSH inequality upon it, let us first do a re-ordering of the Alice's and Bob's bases:
\begin{equation}
    \begin{aligned}
        &{\cal H}_A:\quad  \{\ket{m_0},\ket{ n_0},\ket{ s_1},...,\ket{ s_{d_A-2}}\}_{ s_i\neq m_0, n_0},\\ \bigskip
        &{\cal H}_B:\quad \{\ket{ p_0},\ket{ q_0},\ket{ t_1},...,\ket{ t_{d_B-2}}\}_{ t_j\neq p_0, q_0},
\label{ordering}
    \end{aligned}
\end{equation}
so the two first eigenstates correspond to the indices appearing in the anchor entry. The ordering of the rest of the basis states is irrelevant.

Now, for the observables $A_{1,2}$, $B_{1,2}$ involved in the CHSH inequality (\ref{eq:sum_chsh_popescu}), we take the generic form:
\begin{equation}
    A_i = \begin{pmatrix}
    \vec{a}_i\cdot\vec{\sigma} & \mathbf{0} \\
    \mathbf{0} & \mathds{1}_{d_A - 2} 
    \end{pmatrix} = \vec{a}_i\cdot\vec{\sigma} \oplus \mathds{1}_{d_A - 2},
    \ \ \
    B_i = \begin{pmatrix}
    \vec{b}_i\cdot\vec{\sigma} & \mathbf{0} \\
    \mathbf{0} & \mathds{1}_{d_B - 2} 
    \end{pmatrix}  = \vec{b}_i\cdot\vec{\sigma} \oplus \mathds{1}_{d_B - 2}.
    \label{eq:a_b_popescu}
\end{equation}
Here $\vec \sigma=\left(\sigma_x,\sigma_y,\sigma_z\right)$, i.e. the vector of the 
standard Pauli matrices,
 $\vec a_i$, $\vec b_i$ are unitary 3-vectors
and  $\mathds{1}_d$ is the $d\times d$ identity matrix. This construction is reminiscent of that given by Popescu and Rohrlich in \cite{Popescu:2002won}. Following this reference, a good choice for the $\vec{a}_1$ and $\vec{a}_2$ vectors is
\begin{equation}\label{eq:vec_a_bp}
    \vec{a}_1 = (0,0,1), \quad \vec{a}_2 = (1,0,0).
\end{equation}
Besides, we take $\vec b_1$ and $\vec b_2$ as unitary vectors with opposite polar angle:
\begin{equation}
    \vec b_1=\left(\sin{\theta}\cos{\varphi},\sin{\theta}\sin{\varphi},\cos{\theta}\right),\quad \vec b_2=\left(-\sin{\theta}\cos{\varphi},-\sin{\theta}\sin{\varphi},\cos{\theta}\right),
    \label{eq:vec_b_bp}
\end{equation}
with $\theta\in[0,\pi]$ and $\varphi\in[-\pi,\pi]$. Under this choice and after some algebra (see App.~\ref{app:CHSH} for details), the expression of the expectation value of the CHSH operator, $F(\rho)$, reads
\begin{equation}
\begin{aligned}
    F(\rho)  
     = 2\left[\langle{\cal O}_0\rangle_{\rho}+\sin{\theta}\cos{\varphi}\langle{\cal O}_x\rangle_{\rho}+\sin{\theta}\sin{\varphi}\langle{\cal O}_y\rangle_{\rho}+\cos{\theta}\langle{\cal O}_z\rangle_{\rho}\right],
    \label{eq:F_rho_fin}
\end{aligned}
\end{equation}
where we have introduced 4 new observables defined by:
\begin{equation}
\begin{aligned}
    {\cal O}_0=\left(\sigma_z\oplus \mathds{1}_{d_A-2}\right)\otimes\left(\mathds{O}_2\oplus \mathds{1}_{d_B-2}\right),\quad {\cal O}_z=\left(\sigma_z\oplus \mathds{1}_{d_A-2}\right)\otimes\left(\sigma_z\oplus \mathds{O}_{d_B-2}\right),\\
    {\cal O}_x=\left(\sigma_x\oplus \mathds{1}_{d_A-2}\right)\otimes\left(\sigma_x\oplus \mathds{O}_{d_B-2}\right),\quad {\cal O}_y=\left(\sigma_x\oplus \mathds{1}_{d_A-2}\right)\otimes\left(\sigma_y\oplus \mathds{O}_{d_B-2}\right),
    \label{eq:O_operators}
\end{aligned}
\end{equation}
with $\mathds{O}_d$ the $d\times d$ null matrix. Note that the three last terms of \eqref{eq:F_rho_fin} are simply the dot product $ \vec {b}_1\cdot \langle{\vec{\cal O}}\rangle_{\rho}$ with $\langle{\vec{\cal O}}\rangle_{\rho} = \left(\langle{\cal O}_x\rangle_{\rho},\langle{\cal O}_y\rangle_{\rho},\langle{\cal O}_z\rangle_{\rho}\right)$. The maximum of this expression is obtained when 
$\vec b_1$ is aligned with $\langle{\vec{\cal O}}\rangle_{\rho}$, and it is equal to the modulus of this vector. Hence
\begin{equation}        
    \max_{\theta,\varphi}\lbrace F(\rho)\rbrace=2\left[\langle{\cal O}_0\rangle_{\rho}+\sqrt{\langle{\cal O}_x\rangle_{\rho}^2+\langle{\cal O}_y\rangle_{\rho}^2+\langle{\cal O}_z\rangle_{\rho}^2}\right],
    \label{eq:CHSHMax}
\end{equation}
with $\langle{\cal O}_i\rangle_{\rho} = \Tr{{\cal O}_i\rho}$. It remains to evaluate the traces appearing in \eqref{eq:CHSHMax} knowing the block structure of the $\mathcal{O}_i$ operators. The explicit computation is done in App.~\ref{app:CHSH}, yielding the final result 
\begin{equation}
\begin{aligned}
    \langle{\cal O}_0\rangle_{\rho}&=\sum^{d_B-2}_{j = 1}\left[\rho_{\left( m_0  t_j\right)\left( m_0  t_j\right)}-\rho_{\left( n_0  t_j\right)\left( n_0  t_j\right)}+\sum^{d_A-2}_{i = 1} \rho_{\left( s_i  t_j\right)\left( s_i  t_j\right)} \right],\\
    \langle{\cal O}_z\rangle_{\rho}&= \left[\rho_{\left( m_0  p_0\right)\left( m_0  p_0\right)}-\rho_{\left( m_0  q_0\right)\left( m_0  q_0\right)}\right]+\left[\rho_{\left( n_0  q_0\right)\left( n_0  q_0\right)}-\rho_{\left( n_0  p_0\right)\left( n_0  p_0\right)}\right]
    +\sum^{d_A-2}_{i = 1}\left[\rho_{\left( s_i  p_0\right)\left( s_i  p_0\right)}-\rho_{\left( s_i  q_0\right)\left( s_i  q_0\right)}\right],\\
    \langle{\cal O}_x\rangle_{\rho}&=2 \left[\Re{\rho_{\left( m_0  p_0\right)\left( n_0  q_0\right)}}+\Re{\rho_{\left( n_0  p_0\right)\left( m_0  q_0\right)}}
    +\sum^{d_A-2}_{i = 1}\Re{\rho_{\left( s_i  p_0\right)\left( s_i  q_0\right)}}\right],\\
   \langle{\cal O}_y\rangle_{\rho}&=-2 \left[\Im{\rho_{\left( m_0  p_0\right)\left( n_0  q_0\right)}}+\Im{\rho_{\left( n_0  p_0\right)\left( m_0  q_0\right)}}
    +\sum^{d_A-2}_{i = 1}\Im{\rho_{\left( s_i  p_0\right)\left( s_i  q_0\right)}}\right].
\end{aligned}
\label{eq:Traces_result}
\end{equation}
Here, $\Re{z},\Im{z}$ denote the real and imaginary part of the complex number $z$. 
Now, we can take advantage of the conditions of Eq.~(\ref{eq:SymAssump}) to simplify Eq.~(\ref{eq:Traces_result}). We start with $\langle{\cal O}_x\rangle_\rho$ and $\langle{\cal O}_y\rangle_\rho$. Since $M_0\neq N_0$, we have 
$M_0+Q_0\neq J$, which implies
\begin{equation}
\begin{aligned}
    \rho_{\left( n_0  p_0\right)\left( m_0  q_0\right)}= 0 .
\end{aligned}
\end{equation}
Similarly, since $ s_i\neq n_0$ and $N_0$ is non-degenerate,
\begin{equation}
\begin{aligned}
       \rho_{\left( s_i  p_0\right)\left( s_i  q_0\right)}= 0\quad\quad \forall  s_i .
\end{aligned}
\end{equation}
Then, the sum $\langle{\cal O}_x\rangle_{\rho}^2+\langle{\cal O}_y\rangle_{\rho}^2$ in Eq.~(\ref{eq:CHSHMax}) simply becomes
\begin{equation}
    \langle{\cal O}_x\rangle_{\rho}^2+\langle{\cal O}_y\rangle_{\rho}^2=4\left[\Re{\rho_{\left( m_0  p_0\right)\left( n_0  q_0\right)}}^2+\Im{\rho_{\left( m_0  p_0\right)\left( n_0  q_0\right)}}^2 \right]=4\abs{\rho_{\left( m_0  p_0\right)\left( n_0  q_0\right)}}^2.
    \label{OxOy}
\end{equation}
Analogously, regarding the terms in $\langle{\cal O}_0\rangle_{\rho}$ and $\langle{\cal O}_z\rangle_{\rho}$ we further use that $Q_0$ is non-degenerate and thus 
\begin{equation}
\begin{aligned}
 \rho_{\left( m_0  q_0\right)\left( m_0  q_0\right)}&=\rho_{\left( n_0  p_0\right)\left( n_0  p_0\right)}=0,
 \\
    \rho_{\left( n_0  t_j\right)\left( n_0  t_j\right)}&=0\quad\quad\quad\quad\quad\quad \forall t_j 
    ,\\
    \rho_{\left( s_i  q_0\right)\left( s_i  q_0\right)}&=0\quad\quad\quad\quad\quad\quad \forall  s_i .
\end{aligned}
\end{equation}
Then $\langle{\cal O}_0\rangle_{\rho}$ and $\langle{\cal O}_z\rangle_{\rho}$ become
\begin{equation}
\begin{aligned}
    \langle{\cal O}_0\rangle_{\rho}&=\sum^{d_B-2}_{j = 1}\left[\rho_{\left( m_0  t_j\right)\left( m_0  t_j\right)}+\sum^{d_A-2}_{i = 1} \rho_{\left( s_i  t_j\right)\left( s_i  t_j\right)} \right],\\
    \langle{\cal O}_z\rangle_{\rho}&= \rho_{\left( m_0  p_0\right)\left( m_0  p_0\right)}+\rho_{\left( n_0  q_0\right)\left( n_0  q_0\right)}
    +\sum^{d_A-2}_{i = 1}\rho_{\left( s_i  p_0\right)\left( s_i  p_0\right)}.
\end{aligned}
\label{eq:Traces_result_diag}
\end{equation}
Note that the terms in expressions (\ref{eq:Traces_result_diag}) are the (possibly) non-vanishing diagonal elements of $\rho$, all entering with positive sign. Therefore
\begin{equation}
    \langle{\cal O}_0\rangle_{\rho}+\langle{\cal O}_z\rangle_{\rho}=\Tr{\rho}=1\implies \langle{\cal O}_0\rangle_{\rho}=1-\langle{\cal O}_z\rangle_{\rho}.
    \label{O0}
\end{equation}
Also  both $\langle{\cal O}_0\rangle_{\rho}\geq 0$ and $\langle{\cal O}_z\rangle_{\rho}\geq 0$ since the the diagonal terms of a density matrix are nonnegative.
Plugging the value of $\langle{\cal O}_0\rangle_{\rho}$ and  $\langle{\cal O}_x\rangle_{\rho}^2+\langle{\cal O}_y \rangle_{\rho}^2$ (Eqs.~(\ref{OxOy}, \ref{O0}))
in Eq.~\eqref{eq:CHSHMax}, we get
\begin{equation}\label{eq:pop_end_dem}
    \max_{\theta,\varphi}\lbrace F(\rho)\rbrace=2\left[1+\sqrt{4\abs{\rho_{\left( m_0  p_0\right)\left( n_0  q_0\right)}}^2+\langle{\cal O}_z\rangle_{\rho}^2}-\langle{\cal O}_z\rangle_{\rho}\right],
   \end{equation} 
and thus  
\begin{equation}
    \max_{\theta,\varphi}\lbrace F(\rho)\rbrace>2\iff\rho_{\left( m_0  p_0\right)\left( n_0  q_0\right)}\neq0.
\end{equation}
In conclusion, we have proven that if a magnitude $\hat J=\hat J_A + \hat J_B$ is well-defined with value $J$ in a bipartite physical system described by a density matrix $\rho$, then the presence of an ``anchor entry", i.e. a non-vanishing off-diagonal entry of $\rho$ fulfilling conditions \eqref{eq:SymAssump}, implies the violation of a CHSH inequality. The previous result provides a sufficient condition for Bell non-locality which is extremely easy to check. Note that the non-degeneracy condition of Eq.\eqref{eq:SymAssump} is required only for two of the eigenvalues involved in the anchor entry. 

Besides, if all the relevant\footnote{By ``relevant'', we mean those eigenvalues, $(M,P)$, fulfilling $M+P = J$.} eigenvalues of $\hat J_A, \hat J_B$ are non-degenerate, then all the off-diagonal entries of $\rho$ fulfill the conditions (\ref{eq:SymAssump}), as discussed in the previous section. In that case, the presence of a non-vanishing off-diagonal entry of $\rho$ is a necessary and sufficient condition both for entanglement and Bell-violation. This case has physical relevance. E.g. for the $ZZ$ system produced by the decay of the Higgs boson, it is enough to experimentally show that one off-diagonal element of $\rho$ is different from zero to prove both entanglement and CHSH violation. This will be discussed in further detail in the next section. 

\section{Application to $H\rightarrow ZZ$}\label{sec:Bell_higgs}

One of the simplest and most physically relevant cases where there is an additive observable with well-defined value is the decay of one particle into two others, as e.g. in many Higgs and meson decays. In particular, one can consider the decay of a Higgs boson into two fermions (e.g. two $\tau$'s), or  two vector bosons. For the sake of concreteness, we will focus here on the $H\rightarrow ZZ$ decay.

Since the two $Z$ bosons (Alice and Bob) arise from a spinless particle, the spin component along the momentum direction in the center of mass (C.M.) frame, taken for convenience as $J_z$, is vanishing for the joint system. Hence, we take as our additive observable  $\hat J_z=\hat J_z^{(A)}+\hat J_z^{(B)}$, 
which has the well defined value $J_z=0$. Since the eigenvalues of $J_z^{(A)},J_z^{(B)}$ are non-degenerate, the presence of a non-vanishing off-diagonal element in the spin density matrix of the $ZZ$ system implies both that the state is entangled (see section \ref{subsec:summary_ent}) and that there exists a CHSH inequality which is violated (see section \ref{sec:CHSHviol}).

More precisely, 
using the familiar $J_z-$basis for each Hilbert space 
\begin{equation}
    \mathcal{H}_A = \mathrm{span}\left\lbrace\ket{1}_A,\ket{0}_A,\ket{-1}_A\right\rbrace, \quad \mathcal{H}_B = \mathrm{span}\left\lbrace\ket{1}_B,\ket{0}_B,\ket{-1}_B\right\rbrace,
\end{equation}
the $J_Z=0$ constraint leads to a density matrix with the following texture
\begin{equation}
    \rho =  
    \begin{pmatrix}
    0 & 0 & 0 & 0 & 0 & 0 & 0 & 0 & 0\\
    0 & 0 & 0 & 0 & 0 & 0 & 0 & 0 & 0\\
    0 & 0 & a_{11} & 0 & a_{12} & 0 & a_{13}& 0 & 0\\   
    0 & 0 & 0 & 0 & 0 & 0 & 0 & 0 & 0\\
    0 & 0 & a^*_{12} & 0 & a_{22} & 0 & a_{23} & 0 & 0\\
    0 & 0 & 0 & 0 & 0 & 0 & 0 & 0 & 0\\
    0 & 0 & a^*_{13} & 0 & a^*_{23} & 0 & a_{33} & 0 & 0\\
    0 & 0 & 0 & 0 & 0 & 0 & 0 & 0 & 0\\
    0 & 0 & 0 & 0 & 0 & 0 & 0 & 0 & 0\\
    \end{pmatrix}.
    \label{eq:dens_mat_a}
\end{equation}
Now, we expand $\rho$ in the basis of irreducible tensor operators, $T^L_M$:
\bea
\rho=\frac{1}{9}\left[
\mathbb{1}_3\otimes \mathbb{1}_3+A^1_{LM}\ T^L_{M}\otimes \mathbb{1}_3 + A^2_{LM}\ \mathbb{1}_3\otimes T^L_{M}
+C_{L_1 M_1 L_2 M_2}\ T^{L_1}_{M_1}\otimes T^{L_2}_{M_2}
\right],
\label{rhoAC}
\eea
where sum over $L= 1, 2$ and $-L\leq M\leq L$ (and similarly over $L_{1,2}$ and $M_{1,2}$) is understood. The explicit form of $T^L_M$ and the physical meaning of the coefficients in front are given in \cite{Aguilar-Saavedra:2022wam}. Then the $a_{ij}$ entries of the density matrix read
\begin{equation}
    {a_{12}} = {a_{23}} = \frac{1}{3}{C_{2,1,2,-1}}, \quad {a_{11}} =a_{33} = \frac{1}{3}\left(\frac{1}{\sqrt{2}}A^{1}_{2,0}+1\right), \quad
    a_{13} =  \frac{1}{3}{C_{2,2,2,-2}}.
    \label{relations}
\end{equation}
Besides, the conservation of parity in the decay imposes $C_{2,2,2-2} = \frac{1}{\sqrt{2}}A^1_{2,0} + 1$. For a discussion of radiative corrections on the previous structure, see ref.\cite{Fabbri_2024}.
From the previous discussion, it follows that for either $C_{2,1,2,-1}\neq 0$ and/or $C_{2,2,2,-2}\neq 0$ the system is entangled {\em and} violates a CHSH inequality. We discuss both aspects in order.

Concerning entanglement, according to the simulations performed in \cite{Aguilar-Saavedra:2022wam}, entanglement in $H\rightarrow ZZ$ could be certified at $3\sigma$ ($10\sigma$) at LHC with luminosity $L=300\ {\rm fb}^{-1}$ ($L=3\ {\rm ab}^{-1}$), see tables \ref{tab:Lum_300}, \ref{tab:Lum_3}. 

Concerning Bell nonlocality, it is worth mentioning that in the above-cited reference and in the subsequent literature (see e.g. \cite{Fabbrichesi:2023cev,Morales:2023gow,Bi:2023uop,Barr:2024djo}) the authors focused on the violation of the so-called CGLMP inequalities \cite{Collins:2002sun}, which are somewhat more involved than the CHSH ones. Generically, the CGLMP inequalities are more powerful than the CHSH ones for systems beyond qubit-qubit, as it is the case at hand. However, in this particular instance we stress that, provided an off-diagonal element of $\rho$ is different from zero, the system is not only entangled, but it also violates a CHSH inequality.

In order to explicitly show this, let us choose an off-diagonal entry of the density matrix, which will play the role of the anchor entry, defined in Eqs. (\ref{elemento}-\ref{anchor2}), say $a_{12}=\rho_{\left(1,-1\right)\left(0,0\right)}$. Following the re-labeling given in Eq.(\ref{ordering}), we re-order the basis states as
\begin{equation}
    \mathcal{H}_A = \mathrm{span}\left\lbrace\ket{1},\ket{0},\ket{-1}\right\rbrace, \ \ \mathcal{H}_B = \mathrm{span}\left\lbrace\ket{-1},\ket{0},\ket{1}\right\rbrace.
\end{equation}
In this basis the density matrix has the following texture
\begin{equation}
    \rho =  
    \begin{pmatrix}
    a_{11} & 0 & 0 & 0 & \boxed{a_{12}} & 0 & 0 & 0 & a_{13}\\
    0 & 0 & 0 & 0 & 0 & 0 & 0 & 0 & 0\\
    0 & 0 & 0 & 0 & 0 & 0 & 0 & 0 & 0\\     
    0 & 0 & 0 & 0 & 0 & 0 & 0 & 0 & 0\\
    a^*_{12} & 0 & 0 & 0 & a_{22} & 0 & 0 & 0 & a_{23}\\
    0 & 0 & 0 & 0 & 0 & 0 & 0 & 0 & 0\\
    0 & 0 & 0 & 0 & 0 & 0 & 0 & 0 & 0\\
    0 & 0 & 0 & 0 & 0 & 0 & 0 & 0 & 0\\
    a^*_{13} & 0 & 0 & 0 & a^*_{23} & 0 & 0 & 0 & a_{33}\\
    \end{pmatrix},
    \label{eq:dens_mat_chsh_3_basetransf_popesc}
\end{equation}
where we have boxed the anchor entry upon which we build the CHSH inequality. Using the prescription (\ref{eq:a_b_popescu}), the $A_{1,2}$, $B_{1,2}$ observables read
\begin{equation}
    A_i = \begin{pmatrix}
    \vec{a}_i\cdot\vec{\sigma} & 0 \\
    0 & 1
    \end{pmatrix},
    \ \ \
    B_j = \begin{pmatrix}
    \vec{b}_j\cdot\vec{\sigma} & 0 \\
    0 & 1 
    \end{pmatrix},
    \label{eq:u_a_b_popescu}
\end{equation}
with $\vec{a}_i, \vec{b}_i$ given by Eqs. (\ref{eq:vec_a_bp}, \ref{eq:vec_b_bp}). Then, upon optimization in the $\theta, \varphi$ angles\footnote{The optimal value is given for $\theta=\arccos\left(\dfrac{a_{11}+a_{22}}{\sqrt{4 \abs{a_{12}}^2+\left(a_{11}+a_{22}\right)^2 }}\right)$ and $\varphi=\text{sgn}\left(\Im{a_{12}}\right)\arccos\left(\dfrac{\Re{a_{12}}}{2\abs{a_{12}}}\right)$, with $\text{sgn}\left(x\right)=1 $ if $x>0$, $\text{sgn}\left(x\right)=0 $ if $x=0$ and $\text{sgn}\left(x\right)=-1$ otherwise.}, the corresponding CHSH inequality $F(\rho)\leq 2$ that is violated is given by
Eqs. (\ref{eq:Traces_result_diag}, \ref{eq:pop_end_dem}).
More precisely,
\begin{equation}
    F(\rho) = 2\left[1+ \sqrt{4 \abs{a_{12}}^2+\left(a_{11}+a_{22}\right)^2 }-\left(a_{11}+a_{22}\right)\right],
    \label{eq:CHSH_d=3}
\end{equation}
which guarantees $F(\rho) > 2$ whenever $a_{12} \neq 0$.

We can construct CHSH inequalities upon $a_{23}$ and $a_{13}$ in a similar way (since $a_{12}=a_{23}$, the former is identical to that associated with $a_{12}$). In an obvious notation,
\begin{equation}
\begin{aligned}
    F(\rho)_{12} = F(\rho)_{23} =&\ 2\left[1+ \sqrt{4 \abs{a_{12}}^2+\left(a_{11}+a_{22}\right)^2 }-\left(a_{11}+a_{22}\right)\right], \\
    F(\rho)_{13} =& \ 2\left[1+ \sqrt{4 \abs{a_{13}}^2+\left(a_{11}+a_{33}\right)^2 }-\left(a_{11}+a_{33}\right)\right]
\end{aligned}
\label{eq:CHSH_Higgs}
\end{equation}
(notice that the latter is obtained from Eq.~(\ref{eq:CHSH_d=3}) by permutation of the 2,3 indices).
We can simplify the above expressions by using the relations (\ref{relations}) and parity conservation:
\begin{equation}
\begin{aligned}
    F(\rho)_{12} = F(\rho)_{23} =& \ 2\left[\sqrt{4 \abs{a_{12}}^2+\left(1-\abs{a_{13}}\right)^2 }+\abs{a_{13}}\right], \\
    F(\rho)_{13} =&\ 2 + 4\abs{a_{13}}\left(\sqrt{2} -1\right).
\end{aligned}
\label{eq:CHSH_Higgs_2}
\end{equation}
Since 
$F(\rho)_{13}-2$ and $\abs{a_{13}}$ are proportional, so are their uncertainties. Therefore, the level of confidence at which $a_{13}$ is different from zero is the same as the one at which $F(\rho)_{13}$ is bigger than 2. 
This is also the case for $F(\rho)_{12}$ vs. $a_{12}$, even though $F(\rho)_{12}$ depends on both, $a_{12}$ and $a_{13}$. The reason is that $F(\rho)_{12}>0$ {\em iff} $a_{12}>0$.

Now, from the simulated data presented in \cite{Aguilar-Saavedra:2022wam}, we can compute $F(\rho)_{12}$ and $F(\rho)_{13}$.  The results are shown in tables \ref{tab:Lum_300} and \ref{tab:Lum_3}.
Clearly, both entanglement and violation of CHSH inequalities in $H\rightarrow ZZ$ could be verified at $\sim 3\sigma$ ($\sim 10\sigma$) at LHC with $L=300\ {\rm fb}^{-1}$ ($L=3\ {\rm ab}^{-1}$), by using $a_{12}$ to signal both phenomena and at $\sim 2\sigma$ ($\sim 5\sigma$) by using $a_{13}$.
This represents an important net improvement with respect to the results obtained in \cite{Aguilar-Saavedra:2022wam} (and subsequent literature), where the sensitivity to the violation of the Bell (CGLMP) inequalities was below $2\sigma$ ($5\sigma$) at LHC with $L=300\ {\rm fb}^{-1}$ ($L=3\ {\rm ab}^{-1}$).

Finally, let us mention that in ref. \cite{Bernal:2023ruk,Bernal:2024xhm} it has been recently shown that for the $H\rightarrow ZZ$ system, the presence of entanglement also implies the violation of a different and stronger CGLMP inequality.

\begin{table}[ht]
\centering
\begin{tabular}{c|cccc|}
\cline{2-5}
                                     & 0                               & $10 \ GeV$                      & $20 \ GeV$                       & $30 \ GeV$                        \\ \hline
\multicolumn{1}{|c|}{$N$}            & $450$                           & $418$                           & $312$                            & $129$                             \\
\multicolumn{1}{|c|}{$a_{12}$}  & $-0.33 \pm 0.10$                & $-0.32 \pm 0.11$                & $-0.35 \pm 0.13$                 & $-0.35 \pm 0.20$                  \\
\multicolumn{1}{|c|}{$a_{13}$} & $0.20 \pm 0.12$                 & $0.21 \pm 0.13$                 & $0.25 \pm 0.14$                  & $0.27 \pm 0.21$                   \\
\multicolumn{1}{|c|}{$F(\rho)_{12}$} & $2.47$  & $2.46$  & $2.55$ & $2.57$ \\
\multicolumn{1}{|c|}{} & $> 2~(3.2\sigma)$ & $> 2~(2.9\sigma)$ & $> 2~(2.7\sigma)$ &  $> 2~(1.7\sigma)$ \\ 

\multicolumn{1}{|c|}{$F(\rho)_{13}$} & $2.33$ & $2.35$ & $2.41$ & $2.45$  \\ 
\multicolumn{1}{|c|}{} & $> 2~(1.7\sigma)$ & $> 2~(1.6\sigma)$ & $> 2~(1.8\sigma)$ &  $> 2~(1.3\sigma)$ \\ \hline \end{tabular}
\caption{Values of the crossed off-diagonal terms $a_{12}$ and $a_{13}$ signaling quantum entanglement, as obtained from 1000 pseudoexperiments with $L = 300 ~\mathrm{fb}^{-1}$ in ref. \cite{Aguilar-Saavedra:2022wam}. The violation of the corresponding CHSH inequalities ($F(\rho)_{12}\leq 0$, $F(\rho)_{13}\leq 0$) is reflected in rows (4,5) and (5,6) respectively, where the central value of $F(\rho)$ and the confidence level at which $F(\rho)>2$, thus signaling Bell-violation.}
    \label{tab:Lum_300}
\end{table}

\begin{table}[ht]
\centering 
\begin{tabular}{c|cccc|}
\cline{2-5}
                                     & 0                               & $10 \ GeV$                      & $20 \ GeV$                       & $30 \ GeV$                        \\ \hline
\multicolumn{1}{|c|}{$N$}            & $4500$                           & $4180$                           & $3120$                            & $1290$                             \\
\multicolumn{1}{|c|}{$a_{12}$}  & $-0.32 \pm 0.03$                & $-0.33 \pm 0.03$                & $-0.35 \pm 0.04$                 & $-0.35 \pm 0.06$                  \\
\multicolumn{1}{|c|}{$a_{13}$} & $0.20 \pm 0.04$                 & $0.21 \pm 0.04$                 & $0.25 \pm 0.05$                  & $0.28 \pm 0.07$                   \\
\multicolumn{1}{|c|}{$F(\rho)_{12}$} & $2.44$  & $2.49$  & $2.54$ & $2.56$ \\
\multicolumn{1}{|c|}{} & $> 2~(9.5\sigma)$ & $> 2~(10.0\sigma)$ & $> 2~(8.7\sigma)$ &  $> 2~(5.5\sigma)$ \\ 
\multicolumn{1}{|c|}{$F(\rho)_{13}$} & $2.33$ & $2.35$ & $2.41$ & $2.46$  \\
\multicolumn{1}{|c|}{} & $> 2~(5.0\sigma)$ & $> 2~(5.3\sigma)$ & $> 2~(5.3\sigma)$ &  $> 2~(4.2\sigma)$ \\ \hline
\end{tabular}
\caption{The same as table \ref{tab:Lum_300}, but for a luminosity
 $L = 3~ \mathrm{ab}^{-1}$.}
    \label{tab:Lum_3}
\end{table}

\newpage
\section{Summary and Conclusions}
\label{sec:conclusions}

For generic bipartite mixed states,  sufficient and necessary conditions to certify both entanglement and nonlocality are not known. In particular, the celebrated Peres-Horodecki criterion provides a general sufficient condition for entanglement, but such criterion is  necessary only in the lowest dimension cases: qubit-qubit and qubit-qutrit. 
In this paper we have shown that there is a quite large and physically relevant class of bipartite systems, for which such certifications are pretty feasible. 

Namely, we have considered systems for which there exists an additive observable, say $\hat J=\hat J_A+\hat J_B$, which has a definite value, say $J$. Examples of this class are the final states of Higgs and meson decays into two particles,  with $\hat J$ being the angular-momentum component along the direction of decay in the C.M. frame; spin chains under a given bipartition, with $\hat J$ being e.g.  the magnetization; and hybrid (discrete-continuous) systems, such as atom-cavity setups, where the role of $\hat{J}$ could be played by the energy or by some parity-like property.

Denoting $M,N,\dots$ the eigenvalues of 
$\hat J_A$; $P,Q,\dots$ the eigenvalues of 
$\hat J_B$; and $\ket{m}, \ket{n}$,$\dots$  $\ket{p}, \ket{q}$,$\dots$ the corresponding eigenstates (there are several ones for each eigenvalue in the case of degeneration); our main results are the following:

\begin{itemize}
\item 
If the density matrix contains a non-vanishing {\em crossed off-diagonal entry}, i.e.
$
\rho_{\left( m  p\right)\left( nq\right)}
\neq 0$ with $M+Q\neq J$,
then the state is necessarily entangled and satisfies the Peres-Horodecki criterion. This condition is extremely easy to check.

\item 
If all the crossed off-diagonal entries are vanishing, 
the Peres-Horodecki criterion is a sufficient {\em and} necessary condition provided that for any pair of eigenvalues $\left(M,Q\right)$ satisfying $M+Q=J$, the degeneracy ${\rm deg}\,M\times {\rm deg}\,Q$ is either
\begin{enumerate}
    \item \hspace{0.3cm} ${\rm deg}\,M\times 1\ $ or $\ 1\times {\rm deg}\,Q$.
    \item 
    \hspace{0.3cm} $2\times2$,~ $2\times3$~ or~ $3\times2$.
\end{enumerate}

\item 
If for all pairs $\left(M,Q\right)$ the degeneracy is of type 1, then $\rho$ is entangled if and only if there exists a nonzero crossed off-diagonal element.
Consequently, the Peres-Horodecki criterion holds as a sufficient {\em and} necessary condition. This is the case of qubit-qudit systems, since all the sectors have non-degenerate Alice's eigenvalue (unless $\hat{J}_A\propto\mathds{1}_2 $, i.e. trivial).

\item 
For the degeneracy of type 2, the presence of crossed off-diagonal elements is a sufficient condition for entanglement, but not necessary. Even in its absence, the entanglement can arise from the degenerate subspaces. Still, the Peres-Horodecki criterion is sufficient and necessary.

\item The presence of an {\em anchor element} in the density matrix implies the violation of a CHSH inequality. An anchor element, $\rho_{\left( m_0 p_0\right)\left(n_0 q_0\right)}\neq 0$, is by definition a crossed off-diagonal entry (i.e. $M_0+Q_0\neq J$), such that $N_0$ and $Q_0$ are non-degenerate. This provides a sufficient condition for Bell non-locality which is extremely easy to check. 

\end{itemize}

From the previous facts it follows that if all the
eigenvalues $(M,P)$, fulfilling $M+P = J$ are non-degenerate, then all the off-diagonal entries of $\rho$ are both crossed off-diagonal and anchor entries.
Then, the presence of a non-vanishing off-diagonal entry of $\rho$ is a necessary and sufficient condition for both  entanglement and Bell-violation. 
E.g. for the $ZZ$ system produced by the decay of the Higgs boson, it is enough to experimentally show that one off-diagonal element of $\rho$ is different from zero to prove entanglement, which was already noticed in \cite{Aguilar-Saavedra:2022wam}, and CHSH violation. Using the simulated data presented in this same reference, we have shown that both entanglement and violation of CHSH inequalities in $H\rightarrow ZZ$ could be verified at $\sim 3\sigma$ ($\sim 10\sigma$) at LHC with $L=300\ {\rm fb}^{-1}$ ($L=3\ {\rm ab}^{-1}$), which
represents an important improvement with respect to previous results.

\section*{Acknowledgements}
We are grateful to Jesús M. Moreno for useful discussions. The authors acknowledge the support of the Spanish Agencia Estatal de Investigacion through the grants ``IFT Centro de Excelencia Severo Ochoa CEX2020-001007-S" and PID2022-142545NB-C22 funded by MCIN/AEI/10.13039/501100011033 and by ERDF. The work of A.B. is supported through the FPI grant PRE2020-095867 funded by MCIN/AEI/10.13039/501100011033. J.F. acknowledges support through the Trinity Research Doctorate Award programme 2024-28. Also, J.F. would like to express his sincere thanks to the IFT for its hospitality, especially to students, professors and supervisors during his Master studies.

\appendix
\section{Optimization for the CHSH inequality violation}\label{app:CHSH}
In this appendix we detail the computations performed to derive the results of section \ref{sec:CHSH}.

We first recall that the observables $A_{1,2}$, $B_{1,2}$ involved in the CHSH inequality (\ref{eq:sum_chsh_popescu}) were of the form
\begin{equation}
    A_i = \begin{pmatrix}
    \vec{a}_i\cdot\vec{\sigma} & \mathbf{0} \\
    \mathbf{0} & \mathds{1}_{d_A - 2} 
    \end{pmatrix} = \vec{a}_i\cdot\vec{\sigma} \oplus \mathds{1}_{d_A - 2},
    \ \ \
    B_j = \begin{pmatrix}
    \vec{b}_j\cdot\vec{\sigma} & \mathbf{0} \\
    \mathbf{0} & \mathds{1}_{d_B - 2} 
    \end{pmatrix}  = \vec{b}_j\cdot\vec{\sigma} \oplus \mathds{1}_{d_B - 2}.
    \label{eqApp:a_b_popescu}
\end{equation}
Here, $\vec a_i$, $\vec b_i$ are unitary vectors
and  $\mathds{1}_d$ is the $d\times d$ identity matrix. The choice for the unitary vectors $\vec{a}_1$ and $\vec{a}_2$ vectors is
\begin{equation}\label{eqApp:vec_a_bp}
    \vec{a}_1 = (0,0,1), \quad \vec{a}_2 = (1,0,0).
\end{equation}
while $\vec b_1$ and $\vec b_2$ are taken as unitary vectors with opposite polar angle:
\begin{equation}
    \vec b_1=\left(\sin{\theta}\cos{\varphi},\sin{\theta}\sin{\varphi},\cos{\theta}\right),\quad \vec b_2=\left(-\sin{\theta}\cos{\varphi},-\sin{\theta}\sin{\varphi},\cos{\theta}\right).
    \label{eqApp:vec_b_bp}
\end{equation}
Let us now define $\vec{c}_1$ and $\vec{c}_2$ such that
\begin{equation}
    \vec{b}_1 + \vec{b}_2 = 2\cos(\theta)\vec{c}_1, \ \ \ \vec{b}_1 - \vec{b}_2 = 2\sin(\theta)\vec{c}_2.
    \label{eqApp:def_c_cp}
\end{equation}
More explicitly,
\begin{equation}
    \vec{c}_1=\left(0,0,1\right),\quad \vec{c}_2=\left(\cos{\varphi},\sin{\varphi},0\right).
\end{equation}
We are interested in the operators $B_1 + B_2$, $B_1 - B_2$ that appear in the CHSH inequality (\ref{eq:sum_chsh_popescu}). From Eqs.~(\ref{eqApp:a_b_popescu}, \ref{eqApp:def_c_cp}):
\begin{equation}
    B_1 + B_2 = 2\cos(\theta)C_1, \ \ \ 
    B_1 - B_2 = 2\sin(\theta)C_2, \ \ \ 
    \label{eqApp:def_C_CP}
\end{equation}
with
\begin{equation}
    C_1 = \begin{pmatrix}
    \vec{c}_1\cdot\vec{\sigma} & \mathbf{0} \\
    \mathbf{0} & \mathds{1}_{d_B - 2} 
    \end{pmatrix} =  \vec{c}_1\cdot\vec{\sigma} \oplus \mathds{1}_{d_B - 2},
    \ \ \
    C_2 = \begin{pmatrix}
    \vec{c}_2\cdot\vec{\sigma} & \mathbf{0} \\
    \mathbf{0} & \mathds{O}_{d_B - 2} 
    \end{pmatrix} =  \vec{c}_2\cdot\vec{\sigma} \oplus \mathds{O}_{d_B - 2},
    \label{eqApp:CCppopescu}
\end{equation}
where $\mathds{O}_d$ is the $d\times d$ null matrix.

With these definitions, we can expand the tensor products $A_1\otimes\left(B_1+B_2\right)$ and $A_2\otimes\left(B_1-B_2\right)$ that appear in $F(\rho)$, Eq.~(\ref{eq:sum_chsh_popescu}):
\begin{equation}
\begin{aligned}
    A_1\otimes \left(B_1+B_2\right)&=2\left[\left(\sigma_z\oplus \mathds{1}_{d_A-2}\right)\otimes\left(\cos{\theta}\ \sigma_z\oplus \mathds{1}_{d_B-2}\right)\right]\\
    &=2\left[\left(\sigma_z\oplus \mathds{1}_{d_A-2}\right)\otimes\left(\cos{\theta}\ \sigma_z\oplus \mathds{O}_{d_B-2}\right)+\left(\sigma_z\oplus \mathds{1}_{d_A-2}\right)\otimes\left(\mathds{O}_2\oplus \mathds{1}_{d_B-2}\right)\right]\\
    &=2\left[\cos{\theta}\ {\cal O}_z+{\cal O}_0\right],\\
    A_2\otimes \left(B_1-B_2\right)&=2\left[\left(\sigma_x\oplus \mathds{1}_{d_A-2}\right)\otimes\left(\left(\sin{\theta}\cos{\varphi}\ \sigma_x+\sin{\theta}\sin{\varphi}\ \sigma_y\right)\oplus \mathds{O}_{d_B-2}\right)\right]\\
    &=2\left[\left(\sigma_x\oplus \mathds{1}_{d_A-2}\right)\otimes\left(\sin{\theta}\cos{\varphi}\ \sigma_x\oplus \mathds{O}_{d_B-2}\right)+\left(\sigma_x\oplus \mathds{1}_{d_A-2}\right)\otimes\left(\sin{\theta}\sin{\varphi}\ \sigma_y\oplus \mathds{O}_{d_B-2}\right)\right]\\
    &=2\left[\sin{\theta}\cos{\varphi}\ {\cal O}_x+\sin{\theta}\sin{\varphi}\ {\cal O}_y\right],
\end{aligned}
\end{equation}
where we have introduced 4 new observables defined by:
\begin{equation}
\begin{aligned}
    {\cal O}_0=\left(\sigma_z\oplus \mathds{1}_{d_A-2}\right)\otimes\left(\mathds{O}_2\oplus \mathds{1}_{d_B-2}\right),\quad {\cal O}_z=\left(\sigma_z\oplus \mathds{1}_{d_A-2}\right)\otimes\left(\sigma_z\oplus \mathds{O}_{d_B-2}\right),\\
    {\cal O}_x=\left(\sigma_x\oplus \mathds{1}_{d_A-2}\right)\otimes\left(\sigma_x\oplus \mathds{O}_{d_B-2}\right),\quad {\cal O}_y=\left(\sigma_x\oplus \mathds{1}_{d_A-2}\right)\otimes\left(\sigma_y\oplus \mathds{O}_{d_B-2}\right).
\end{aligned}\label{eqApp:O_operators}
\end{equation}
Thus, $F(\rho)$ is given by:
\begin{equation}
\begin{aligned}
    F(\rho)  
     = 2\left[\langle{\cal O}_0\rangle_{\rho}+\sin{\theta}\cos{\varphi}\langle{\cal O}_x\rangle_{\rho}+\sin{\theta}\sin{\varphi}\langle{\cal O}_y\rangle_{\rho}+\cos{\theta}\langle{\cal O}_z\rangle_{\rho}\right].
     \label{eqApp:F_rho_fin}
\end{aligned}
\end{equation}
In order to evaluate the traces in \eqref{eqApp:F_rho_fin} we need more explicit expressions of $\rho$ and ${\cal O}_0, {\cal O}_x, {\cal O}_y, {\cal O}_z$. All of them have dimension $d_Ad_B\times d_Ad_B$.
Here it is very convenient to divide these matrices in blocks in the following way.  The density matrix $\rho$ can be expressed as
\begin{equation}
\rho = \left(\begin{array}{c|c} 
    \tilde{\rho}_{1,1}  &   \tilde{\rho}_{1,2}  \\
    \hline
    \tilde{\rho}_{2,1}  &   \tilde{\rho}_{2,2} 
\end{array}
\right),
\label{eqApp:block_mat}
\end{equation}
where the size of these four blocks is:
\bea
&\tilde{\rho}_{1,1}:&\ \ \ 2d_B\times2d_B,
\nonumber\\
&\tilde{\rho}_{1,2}:&\ \ \ 2d_B\times(d_A -2)d_B,
\nonumber\\
&\tilde{\rho}_{2,1}:&\ \ \ (d_A -2)d_B\times2d_B,
\nonumber\\
&\tilde{\rho}_{2,2}:&\ \ \ (d_A -2)d_B\times(d_A -2)d_B.
\label{sizes}
\eea
More explicitly,
\begin{equation}
    \tilde{\rho}_{1,1} = 
    \begin{pmatrix}
     \rho_{\left( m_0  p_0\right)\left( m_0 p_0\right)} & \rho_{\left( m_0  p_0\right)\left( m_0 q_0\right)} & \rho_{\left( m_0  p_0\right)\left( m_0 t_1\right)} & \hdots & \rho_{\left( m_0  p_0\right)\left( n_0 p_0\right)} & \boxed{\rho_{\left( m_0  p_0\right)\left( n_0 q_0\right)}} & \rho_{\left( m_0  p_0\right)\left( n_0 t_1\right)} &  \hdots \\
     \rho_{\left( m_0  q_0\right)\left( m_0 p_0\right)}  & \rho_{\left( m_0  q_0\right)\left( m_0 q_0\right)} & \rho_{\left( m_0  q_0\right)\left( m_0 t_1\right)} & \hdots & \rho_{\left( m_0  q_0\right)\left( n_0 p_0\right)} & \rho_{\left( m_0  q_0\right)\left( n_0 q_0\right)} & \rho_{\left( m_0  q_0\right)\left( n_0 t_1\right)} & \hdots \\
     \rho_{\left( m_0  t_1\right)\left( m_0 p_0\right)}  & \rho_{\left( m_0  t_1\right)\left( m_0 q_0\right)} & \rho_{\left( m_0  t_1\right)\left( m_0 t_1\right)} & \hdots & \rho_{\left( m_0  t_1\right)\left( n_0 p_0\right)} & \rho_{\left( m_0  t_1\right)\left( n_0 q_0\right)} & \rho_{\left( m_0  t_1\right)\left( n_0 t_1\right)} & \hdots\\
     \vdots  & \vdots & \vdots & \ddots & \vdots & \vdots & \vdots & \ddots \\
     \rho_{\left( n_0  p_0\right)\left( m_0 p_0\right)}  & \rho_{\left( n_0  p_0\right)\left( m_0 q_0\right)} & \rho_{\left( n_0  p_0\right)\left( m_0 t_1\right)} & \hdots & \rho_{\left( n_0  p_0\right)\left( n_0 p_0\right)} & \rho_{\left( n_0  p_0\right)\left( n_0 q_0\right)} & \rho_{\left( n_0  p_0\right)\left( n_0 t_1\right)} & \hdots\\
     \boxed{\rho_{\left( n_0  q_0\right)\left( m_0 p_0\right)}}  & \rho_{\left( n_0  q_0\right)\left( m_0 q_0\right)} & \rho_{\left( n_0  q_0\right)\left( m_0 t_1\right)} & \hdots & \rho_{\left( n_0  q_0\right)\left( n_0 p_0\right)} & \rho_{\left( n_0  q_0\right)\left( n_0 q_0\right)} & \rho_{\left( n_0  q_0\right)\left( n_0 t_1\right)} & \hdots\\
     \rho_{\left( n_0  t_1\right)\left( m_0 p_0\right)}  & \rho_{\left( n_0  t_1\right)\left( m_0 q_0\right)} & \rho_{\left( n_0  t_1\right)\left( m_0 t_1\right)} & \hdots & \rho_{\left( n_0  t_1\right)\left( n_0 p_0\right)} & \rho_{\left( n_0  t_1\right)\left( n_0 q_0\right)} & \rho_{\left( n_0  t_1\right)\left( n_0 t_1\right)} & \hdots\\
     \vdots  & \vdots & \vdots & \ddots & \vdots & \vdots & \vdots & \ddots 
    \end{pmatrix},
    \label{eqApp:rho_tilde}
\end{equation}
where we have boxed the anchor entry (see Eqs.\eqref{elemento},\eqref{eq:SymAssump}) and its conjugate. Besides,
\begin{equation}
    \tilde{\rho}_{1,2} = \begin{pmatrix}
        \rho_{\left( m_0  p_0\right)\left( s_1 p_0\right)} &\rho_{\left( m_0  p_0\right)\left( s_1 q_0\right)} & \rho_{\left( m_0  p_0\right)\left( s_1 t_1\right)} & \hdots \\
        \vdots  & \vdots & \vdots & \vdots \\ 
        \rho_{\left( n_0  t_1\right)\left( s p_0\right)} &\rho_{\left( n_0  t_1\right)\left( s_1 q_0\right)} & \rho_{\left( n_0  t_1\right)\left( s_1 t_1\right)}& \hdots \\
        \vdots  & \vdots & \vdots & \vdots
    \end{pmatrix} = \left(\tilde{\rho}_{2,1}\right)^{\dagger}
    \label{eqApp:Delta}
\end{equation}
and, finally, 
\begin{equation}
    \tilde{\rho}_{2,2} = 
    \begin{pmatrix}
     \rho_{\left( s_1  p_0\right)\left( s_1 p_0\right)} & \rho_{\left( s_1  p_0\right)\left( s_1 q_0\right)} & \rho_{\left( s_1  p_0\right)\left( s_1 t_1\right)} & \hdots &  \rho_{\left( s_1  p_0\right)\left( s_{d_A-2} t_{d_B-2}\right)}\\
     \rho_{\left( s_1  q_0\right)\left( s_1 p_0\right)}  & \rho_{\left( s_1  q_0\right)\left( s_1 q_0\right)} & \rho_{\left( s_1  q_0\right)\left( s_1 t_1\right)} & \hdots & \rho_{\left( s_1  q_0\right)\left( s_{d_A-2} t_{d_B-2}\right)}\\
     \rho_{\left( s_1  t_1\right)\left( s_1 p_0\right)}  & \rho_{\left( s_1  t_1\right)\left( s_1 q_0\right)} & \rho_{\left( s_1  t_1\right)\left( s_1 t_1\right)} & \hdots & \rho_{\left( s_1  t_1\right)\left( s_{d_A-2} t_{d_B-2}\right)}\\
     \vdots  & \vdots & \vdots & \ddots & \vdots \\
     \rho_{\left( s_{d_A-2}  t_{d_B-2}\right)\left( s_1 p_0\right)} & \rho_{\left( s_{d_A-2}  t_{d_B-2}\right)\left( s_1 q_0\right)} & \rho_{\left( s_{d_A-2}  t_{d_B-2}\right)\left( s_1 t_1\right)} & \hdots &  \rho_{\left( s_{d_A-2}  t_{d_B-2}\right)\left( s_{d_A-2} t_{d_B-2}\right)}
    \end{pmatrix}.
    \label{eqApp:rho_D}
\end{equation}
On the other hand, the $\cal O$-matrices defined in Eq.\eqref{eqApp:O_operators} are given by 
\begin{equation}
\mathcal{O}_{0} = 
\left(\begin{array}{cc|ccc} 
     \begin{pmatrix}
             \mathds{O}_2 & \mathbf{0}\\
             \mathbf{0} & \mathds{1}_{d_B-2}\\
             \end{pmatrix} & \mathds{O}_{d_B} & \mathds{O}_{d_B} & \hdots &  \mathds{O}_{d_B} \\[.3cm]
     \mathds{O}_{d_B} & -\begin{pmatrix}
             \mathds{O}_2 & \mathbf{0}\\
             \mathbf{0} & \mathds{1}_{d_B-2}\\
             \end{pmatrix} & \mathds{O}_{d_B} & \hdots & \mathds{O}_{d_B} \\[.5cm]
              \hline
            \rule{0pt}{1.5\normalbaselineskip} \mathds{O}_{d_B} & \mathds{O}_{d_B} & \begin{pmatrix}
             \mathds{O}_2 & \mathbf{0}\\
             \mathbf{0} & \mathds{1}_{d_B-2}\\
             \end{pmatrix} & \hdots & \mathds{O}_{d_B}\\
             \vdots & \vdots & \vdots & \ddots & \vdots\\
             \mathds{O}_{d_B} & \mathds{O}_{d_B} & \mathds{O}_{d_B} & \hdots & \begin{pmatrix}
             \mathds{O}_2 & \mathbf{0}\\
             \mathbf{0} & \mathds{1}_{d_B-2}\\
             \end{pmatrix}
\end{array}\right), 
    \label{eqApp:O_0}
\end{equation}

\begin{equation}
\hspace*{-.3cm} 
\mathcal{O}_{z} = 
\left(\begin{array}{cc|ccc} 
            \begin{pmatrix}
            \sigma_z & \mathbf{0}\\
             \mathbf{0} & \mathds{O}_{d_B-2}\\
             \end{pmatrix} & \mathds{O}_{d_B} & \mathds{O}_{d_B} & \hdots & \mathds{O}_{d_B} \\[.3cm]
     \mathds{O}_{d_B} & -\begin{pmatrix}
             \sigma_z & \mathbf{0}\\
             \mathbf{0} & \mathds{O}_{d_B-2}\\
             \end{pmatrix} & \mathds{O}_{d_B} & \hdots & \mathds{O}_{d_B} \\[.5cm]
              \hline
            \rule{0pt}{1.5\normalbaselineskip} \mathds{O}_{d_B} & \mathds{O}_{d_B} & \begin{pmatrix}
             \sigma_z & \mathbf{0}\\
             \mathbf{0} & \mathds{O}_{d_B-2}\\
             \end{pmatrix} & \hdots & \mathds{O}_{d_B}\\
             \vdots & \vdots & \vdots & \ddots & \vdots\\
             \mathds{O}_{d_B} & \mathds{O}_{d_B} & \mathds{O}_{d_B} & \hdots & \begin{pmatrix}
             \sigma_z & \mathbf{0}\\
             \mathbf{0} & \mathds{O}_{d_B-2}\\
             \end{pmatrix}
\end{array}\right),
    \label{eqApp:O_z}
\end{equation}
\begin{equation}
\mathcal{O}_{x} = 
\left(\begin{array}{cc|ccc} 
             \mathds{O}_{d_B} & \begin{pmatrix}
             \sigma_x & \mathbf{0}\\
             \mathbf{0} & \mathds{O}_{d_B-2}\\ \end{pmatrix}& \mathds{O}_{d_B} & \hdots & \mathds{O}_{d_B} \\[.3cm]
             \begin{pmatrix}
             \sigma_x & \mathbf{0}\\
             \mathbf{0} & \mathds{O}_{d_B-2}\\ \end{pmatrix} & \mathds{O}_{d_B} & \mathds{O}_{d_B} & \hdots & \mathds{O}_{d_B} \\[.5cm]
              \hline
            \rule{0pt}{1.5\normalbaselineskip} \mathds{O}_{d_B} & \mathds{O}_{d_B} & \begin{pmatrix}
             \sigma_x & \mathbf{0}\\
             \mathbf{0} & \mathds{O}_{d_B-2}\\
             \end{pmatrix} & \hdots & \mathds{O}_{d_B}\\
             \vdots & \vdots & \vdots & \ddots & \vdots \\
             \mathds{O}_{d_B} & \mathds{O}_{d_B} & \mathds{O}_{d_B} & \hdots & \begin{pmatrix}
             \sigma_x & \mathbf{0}\\
             \mathbf{0} & \mathds{O}_{d_B-2}\\
             \end{pmatrix}
\end{array}\right),
    \label{eqApp:O_x}
\end{equation}

\begin{equation}
\mathcal{O}_{y} = 
\left(\begin{array}{cc|ccc} 
             \mathds{O}_{d_B} & \begin{pmatrix}
             \sigma_y & \mathbf{0}\\
             \mathbf{0} & \mathds{O}_{d_B-2}\\ \end{pmatrix}& \mathds{O}_{d_B} & \hdots & \mathds{O}_{d_B} \\[.3cm]
             \begin{pmatrix}
             \sigma_y & \mathbf{0}\\
             \mathbf{0} & \mathds{O}_{d_B-2}\\ \end{pmatrix} & \mathds{O}_{d_B} & \mathds{O}_{d_B} & \hdots & \mathds{O}_{d_B} \\[.5cm]
              \hline
            \rule{0pt}{1.5\normalbaselineskip} \mathds{O}_{d_B} & \mathds{O}_{d_B} & \begin{pmatrix}
             \sigma_y & \mathbf{0}\\
             \mathbf{0} & \mathds{O}_{d_B-2}\\
             \end{pmatrix} & \hdots & \mathds{O}_{d_B}\\
             \vdots & \vdots & \vdots & \ddots & \vdots \\
             \mathds{O}_{d_B} & \mathds{O}_{d_B} & \mathds{O}_{d_B} & \hdots & \begin{pmatrix}
             \sigma_y & \mathbf{0}\\
             \mathbf{0} & \mathds{O}_{d_B-2}\\
             \end{pmatrix}
\end{array}\right),
    \label{eqApp:O_y}
\end{equation}
where we have indicated a block division which matches the one made for $\rho$, see Eqs.\eqref{eqApp:block_mat}, \eqref{sizes}. Now it is straightforward to evaluate $\Tr{\rho\mathcal{O}_{i}}$ by performing block product of these matrices. Note in particular that 
the blocks $\tilde{\rho}_{1,2}$ and $\tilde{\rho}_{2,1}$ of Eq.\eqref{eqApp:block_mat} do not contribute to any of the traces since the off-diagonal blocks of all $\mathcal{O}_{i}$ are identically zero. 
On the other hand, the contributions of the blocks $\tilde{\rho}_{1,1}$ and $\tilde{\rho}_{2,2}$ are easy to compute:
\begin{equation}
\begin{aligned}
    \langle{\cal O}_0\rangle_{\rho}&=\sum^{d_B-2}_{j = 1}\left[\rho_{\left( m_0  t_j\right)\left( m_0  t_j\right)}-\rho_{\left( n_0  t_j\right)\left( n_0  t_j\right)}+\sum^{d_A-2}_{i = 1} \rho_{\left( s_i  t_j\right)\left( s_i  t_j\right)} \right],\\
    \langle{\cal O}_z\rangle_{\rho}&= \left[\rho_{\left( m_0  p_0\right)\left( m_0  p_0\right)}-\rho_{\left( m_0  q_0\right)\left( m_0  q_0\right)}\right]+\left[\rho_{\left( n_0  q_0\right)\left( n_0  q_0\right)}-\rho_{\left( n_0  p_0\right)\left( n_0  p_0\right)}\right]
    +\sum^{d_A-2}_{i = 1}\left[\rho_{\left( s_i  p_0\right)\left( s_i  p_0\right)}-\rho_{\left( s_i  q_0\right)\left( s_i  q_0\right)}\right],\\
    \langle{\cal O}_x\rangle_{\rho}&=2 \left[\Re{\rho_{\left( m_0  p_0\right)\left( n_0  q_0\right)}}+\Re{\rho_{\left( n_0  p_0\right)\left( m_0  q_0\right)}}
    +\sum^{d_A-2}_{i = 1}\Re{\rho_{\left( s_i  p_0\right)\left( s_i  q_0\right)}}\right],\\
   \langle{\cal O}_y\rangle_{\rho}&=-2 \left[\Im{\rho_{\left( m_0  p_0\right)\left( n_0  q_0\right)}}+\Im{\rho_{\left( n_0  p_0\right)\left( m_0  q_0\right)}}
    +\sum^{d_A-2}_{i = 1}\Im{\rho_{\left( s_i  p_0\right)\left( s_i  q_0\right)}}\right].
\end{aligned}
\end{equation}

 \newpage
\bibliographystyle{style2.bst}    
%
\bibliography{references}	 

\end{document}